\documentclass[aps,prd,reprint,nofootinbib,superscriptaddress]{revtex4-2}
\usepackage{amsmath,amssymb}
\usepackage{graphicx}
\usepackage{booktabs}
\graphicspath{{figures/}}

\begin{document}

\title{Greybody Factors, Absorption Cross Sections and Hawking Radiation of Holonomy-Corrected Schwarzschild Black Holes}

\author{Bekir Can L{\"u}tf{\"u}o{\u{g}}lu}
\email{bekir.lutfuoglu@uhk.cz}
\affiliation{Department of Physics, Faculty of Science, University of Hradec Kr{\'a}lov{\'e}, Rokitansk{\'e}ho 62/26, 500 03 Hradec Kr{\'a}lov{\'e}, Czech Republic}

\author{Javlon~Rayimbaev}
\email{javlon@astrin.uz}
\affiliation{Kimyo International University in Tashkent, Shota Rustaveli Street 156, Tashkent 100121, Uzbekistan}

\author{Bekzod Rahmatov}
\email{rahmatovbekzod@samdu.uz}
\affiliation{New Uzbekistan University, Movarounnahr Str. 1, Tashkent 100000, Uzbekistan}




\author{Saidmuhammad Ahmedov} \email{saidmuhammadaxmed@gmail.com} \affiliation{National University of Uzbekistan, Tashkent 100174, Uzbekistan}

\author{Nuriddin Kurbonov}
\email{n.kurbonov@newuu.uz}
\affiliation{Ulugh Beg Astronomical Institute, Astronomy St. 33, Tashkent 100052, Uzbekistan}
\affiliation{University of Tashkent for Applied Sciences, Gavhar Str. 1, Tashkent 700127, Uzbekistan}
\affiliation{Tashkent State Technical University, Tashkent 100095, Uzbekistan}

\date{\today}

\begin{abstract}
We study greybody factors, absorption cross sections and Hawking energy-emission rates for minimally coupled massless scalar, electromagnetic and massless Dirac test fields on the loop-quantum-gravity-inspired holonomy-corrected Schwarzschild black hole. The geometry is controlled by a dimensionless holonomy parameter, and the radial wave equations are solved by direct numerical integration with first- and sixth-order WKB estimates as complementary checks. The scalar, electromagnetic and Dirac channels respond differently: the dominant scalar mode becomes more transparent, the electromagnetic threshold shifts slightly upward, and the dominant Dirac mode is only mildly modified. The scalar absorption cross section retains the universal low-frequency limit, the electromagnetic cross section changes mainly in the infrared, and the Dirac cross section develops a strongly suppressed low-frequency tail. Since the Hawking temperature falls monotonically, thermal suppression dominates the radiative output. Thus the holonomy correction enhances low-lying scalar transmission but suppresses Hawking radiation overall, with the electromagnetic sector most strongly quenched and the fermionic sector dominant once $\alpha$ is appreciable.
\end{abstract}

\maketitle

\section{Introduction}

Black holes do not radiate as perfect black bodies: Hawking radiation is thermal at the horizon, but curvature scattering filters the flux observed at infinity. The corresponding transmission coefficients, or greybody factors, determine partial absorption probabilities, total absorption cross sections and frequency-resolved Hawking spectra \cite{Hawking1975,Page1976,Unruh1976,DasGibbonsMathur1997,Kanti2004,KantiWinstanley2015}. They therefore provide a clean bridge between geometry, wave propagation and black-hole thermodynamics, and have recently been used as complementary probes of quasinormal-mode spectra \cite{Konoplya:2024lir,Konoplya:2024vuj}.

The present paper focuses on a loop-quantum-gravity-inspired, holonomy-corrected Schwarzschild geometry. Rather than choosing an arbitrary regular lapse function, the effective theory starts from the canonical spherically symmetric sector of general relativity in radial Ashtekar--Barbero variables and modifies the extrinsic-curvature dependence by holonomy functions that mimic finite parallel transports. This modification is combined with a suitable recombination of the Hamiltonian and diffeomorphism constraints, keeping the constraint algebra first class and preserving a spacetime interpretation \cite{AlonsoBardajiBrizuela2021EPJC,AlonsoBardajiBrizuela2021PRD,AlonsoBardajiBrizuelaVera2022PLB,AlonsoBardajiBrizuelaVera2022PRD}.

In the covariant effective description developed by Alonso-Bardaji, Brizuela and Vera, these anomaly-free holonomy corrections make the spherically symmetric vacuum spacetime regular and replace the classical singularity by a smooth minimal-radius region \cite{AlonsoBardajiBrizuelaVera2022PLB,AlonsoBardajiBrizuelaVera2022PRD}. The corresponding global structure is that of a black bounce connecting two asymptotically flat regions. In the exterior static patch relevant for scattering, the solution is described by the Schwarzschild horizon radius and by one additional dimensionless holonomy parameter. Thus the geometry remains simple enough for a clean greybody-factor calculation, while still carrying a direct imprint of the underlying quantum-gravity-inspired deformation.

For clarity, the word ``holonomy'' in this paper refers to the polymerization of the angular extrinsic-curvature variable in the symmetry-reduced gravitational phase space, not to an additional matter charge. The associated parameter controls the location of the minimal two-sphere that replaces the Schwarzschild singularity. In the notation used below, this scale is denoted by $r_0$, the outer horizon by $r_h$, and the dimensionless deformation strength by $\alpha=r_0/r_h$. The limit $\alpha=0$ is exactly Schwarzschild; increasing $\alpha$ leaves the asymptotic mass and horizon radius fixed in our comparison but changes the interior completion and the radial part of the exterior metric.

The massless scalar quasinormal spectrum of this spacetime was first studied by Moreira, Lima Junior, Crispino and Herdeiro \cite{MoreiraLimaJuniorCrispinoHerdeiro2023}, and a broader overtone analysis for scalar, electromagnetic and Dirac perturbations was later given by Bolokhov \cite{Bolokhov:2023bwm}. Related work also derived rigorous bounds for greybody factors in the same effective geometry \cite{RinconOvgunPantig2024}. Dedicated greybody and Hawking-flux computations in higher-derivative and four-dimensional Einstein--Gauss--Bonnet settings also show that transmission factors can be essential for interpreting departures from Schwarzschild emission \cite{Konoplya:2019ppy,Konoplya:2020cbv}. In higher-dimensional Gauss--Bonnet gravity, the same interplay can be even more pronounced: the GB coupling suppresses tensor-graviton emission through both rapid cooling and a strong decrease of the tensor greybody factor, leading to much longer evaporation times \cite{Konoplya:2010vz}. Similar combinations of quasinormal ringing, greybody factors and thermal diagnostics have been considered for black holes in galactic environments, regular black holes with scalar hair, generalized Rastall gravity, quantum-corrected geometries and scalarized metrics \cite{Konoplya:2021ube,Konoplya:2023ppx,Karmakar:2023cwg,Skvortsova:2024msa,Lutfuoglu:2026gey,Konoplya:2025hgp}. Scalar, Dirac and gravitational greybody factors in Euler--Heisenberg electrodynamics and asymptotically safe gravity provide further examples where the field content and the modified background both enter the transmission problem \cite{Malik:2025erb,Lutfuoglu:2025ohb}. What is still missing, however, is a dedicated computation of the actual transmission coefficients, the corresponding absorption cross sections and the Hawking energy-emission rates for test scalar, electromagnetic and Dirac fields.

That is the aim of this manuscript. We deliberately keep the analysis simple and focused. First, we work with test fields on a fixed background, so the geometry is not allowed to backreact to the emitted radiation. Second, we concentrate on the minimally coupled massless scalar, Maxwell and massless Dirac fields, because all three reduce to one-dimensional scattering problems with transparent physical interpretation. Third, we use direct numerical integration as the primary method for extracting greybody factors. First- and sixth-order WKB estimates are included only as complementary barrier-top checks, not as the main engine of the calculation. This is important here: the WKB approximation is useful near the top of a single smooth barrier, but the observables of real interest---especially absorption and Hawking emission---must be built from numerically determined transmission probabilities.

The main physical picture turns out to be sharp. The holonomy correction affects scalar, electromagnetic and Dirac propagation differently. The dominant scalar mode becomes easier to transmit, the electromagnetic channel is mildly pushed toward higher threshold frequencies, and the dominant Dirac mode remains almost threshold-invariant. But the same correction lowers the Hawking temperature as $\sqrt{1-\alpha}$, so thermodynamics ultimately wins over transmission when one considers the emitted power. The result is a black hole that is locally more transparent to the scalar $s$-wave, almost neutral in its dominant fermionic threshold, yet globally dimmer as a Hawking emitter.

The paper is organized as follows. Section~\ref{sec:geometry} introduces the geometry, the tortoise coordinate and the master equations for the scalar, electromagnetic and Dirac fields. Section~\ref{sec:methods} explains the numerical integration scheme and the complementary WKB estimate, and gives the formulas for absorption cross sections and Hawking radiation. Section~\ref{sec:gbf} discusses the greybody factors and  absorption cross sections. Section~\ref{sec:hawking} analyzes the Hawking spectra, including a Page-style aggregate $3\,\mathrm{Dirac}+\gamma$, and the integrated emitted powers. Section~\ref{sec:conclusion} summarizes the overall picture.

\section{Holonomy-Corrected Geometry and Test-Field Equations}
\label{sec:geometry}

\textbf{Metric and Hawking temperature.}
Before specializing to the static exterior metric, it is useful to recall the minimal structure of the effective gravity model. The canonical variables are functions of a radial coordinate $x$ and of time. The pair $(E^x,K_x)$ describes the radial triad component and its conjugate extrinsic-curvature component, while $(E^\varphi,K_\varphi)$ describes the angular part. In particular, $E^x$ fixes the areal radius through $r=\sqrt{E^x}$, and $E^\varphi$ controls the radial metric coefficient. The lapse $N$ and shift $N^x$ are Lagrange multipliers imposing the Hamiltonian and radial diffeomorphism constraints, respectively.

The characteristic loop-inspired input is the replacement of the angular extrinsic-curvature dependence by a bounded polymerized function,
\begin{equation}
\label{eq:holonomy_map}
b_\lambda(K_\varphi)=\frac{\sin(\lambda K_\varphi)}{\lambda},
\quad
\lim_{\lambda\to0}b_\lambda(K_\varphi)=K_\varphi,
\end{equation}
where $\lambda$ is the dimensionless polymerization parameter. This equation summarizes the classical limit: when $\lambda$ is sent to zero, the usual connection/extrinsic-curvature dependence of spherical general relativity is recovered. For finite $\lambda$, the sine function bounds the effective curvature contribution and is responsible for the singularity-resolving behavior of the model.

The important consistency requirement is not only the replacement \eqref{eq:holonomy_map}, but also anomaly freedom. We denote by $D[\xi]$ the radial diffeomorphism constraint smeared with a shift-like test function $\xi(x)$, and by $H[N]$ the Hamiltonian constraint smeared with the lapse $N(x)$. After an appropriate recombination of the constraints, the deformed generators still form a first-class algebra of the schematic form
\begin{align}
\{D[\xi],D[\eta]\}&=D[\xi\eta'-\xi'\eta],
\label{eq:constraint_algebra_1}\\
\{D[\xi],H[N]\}&=H[\xi N'],
\label{eq:constraint_algebra_2}\\
\{H[N],H[\widetilde N]\}&=D\!\left[\beta_\lambda(E,K)\left(N\widetilde N'-N'\widetilde N\right)\right].
\label{eq:constraint_algebra_3}
\end{align}
Here $\eta(x)$ and $\widetilde N(x)$ are additional smearing functions, a prime denotes differentiation with respect to $x$, and $\beta_\lambda(E,K)$ is a holonomy-deformed structure function built from the canonical variables. Its detailed expression is not needed for the scattering calculation; what matters is that it reduces to the classical spherical structure function when $\lambda\to0$. This closure is what allows the effective solution to be interpreted as a spacetime geometry rather than only as a gauge-fixed ansatz.

For the static vacuum black-hole solution constructed in Refs.~\cite{AlonsoBardajiBrizuelaVera2022PLB,AlonsoBardajiBrizuelaVera2022PRD}, the polymerization parameter fixes the minimal-radius scale according to
\begin{equation}
\label{eq:r0_lambda}
r_0=\frac{2M\lambda^2}{1+\lambda^2},
\quad
r_h=2M,
\quad
\alpha\equiv\frac{r_0}{r_h}=\frac{\lambda^2}{1+\lambda^2}.
\end{equation}
The symbol $M$ is the mass parameter measured by the asymptotic Schwarzschild term, $r_h$ is the outer event-horizon radius, $r_0$ is the minimum areal radius reached by the effective geometry, and $\alpha$ is the dimensionless holonomy parameter used in all plots and tables. Thus the Schwarzschild limit corresponds to $\lambda=0$ and $\alpha=0$, while increasing $\lambda$ pushes the minimal-radius surface toward the event horizon. The formal limit $\lambda\to\infty$ gives $\alpha\to1$, which is why the exterior family is taken with $0\le \alpha<1$.

With these definitions, the exterior chart used in Refs.~\cite{AlonsoBardajiBrizuelaVera2022PLB,AlonsoBardajiBrizuelaVera2022PRD,MoreiraLimaJuniorCrispinoHerdeiro2023} can be written in the form
\begin{equation}
\begin{split}
\label{eq:metric}
 ds^2&=-F(r)\,dt^2+\frac{dr^2}{G(r)}+r^2 d\Omega_2^2,\\
F(r)&=1-\frac{r_h}{r},\\
G(r)&=\left(1-\frac{\alpha r_h}{r}\right)
\left(1-\frac{r_h}{r}\right),
\quad 0\le \alpha<1.
\end{split}
\end{equation}
Here $r_h=2M$ is the event-horizon radius and $\alpha=r_0/r_h$ measures the strength of the holonomy correction, with $r_0$ the minimal-radius scale introduced above. The Schwarzschild limit is recovered at $\alpha=0$, while $\alpha\to1^-$ approaches the extremal end of the holonomy-corrected family \cite{MoreiraLimaJuniorCrispinoHerdeiro2023,Bolokhov:2023bwm}. Throughout the numerical analysis we set $r_h=1$, so every frequency is measured in the dimensionless combination $\omega r_h$ and every cross section in units of $r_h^2$.

For a metric of the form \eqref{eq:metric}, the surface gravity is
$
\kappa=\frac{1}{2}\sqrt{F'(r_h)G'(r_h)},
$
which gives the Hawking temperature
\begin{equation}
\label{eq:temperature}
T_H=\frac{\kappa}{2\pi}=\frac{\sqrt{1-\alpha}}{4\pi r_h}.
\end{equation}
Thus the holonomy correction cools the black hole monotonically. This simple square-root suppression will later dominate the emitted flux.
The tortoise coordinate is defined by
\begin{equation}
\label{eq:tortoise}
\frac{dr_*}{dr}=\frac{1}{\sqrt{F(r)G(r)}}=
\frac{1}{\left(1-r_h/r\right)\sqrt{1-\alpha r_h/r}}.
\end{equation}
As usual, $r_*\to-\infty$ at the horizon and $r_*\sim r$ at large radius. With this coordinate, the reduced radial functions for all three test fields considered below can be written in the common Schr\"odinger-like form
\begin{equation}
\label{eq:master_common}
\frac{d^2\Psi_a}{dr_*^2}+\left[\omega^2-V_a(r)\right]\Psi_a=0,
\end{equation}
where $a$ labels the field and angular mode. The field dependence enters through the effective potential $V_a(r)$.

\textbf{Scalar field.}
For a minimally coupled massless scalar field satisfying
$\Box\Phi=0,$
we separate variables as \cite{Carter:1968ks,Konoplya:2018arm}
\begin{equation}
\Phi(t,r,\theta,\varphi)=\sum_{\ell,m}\frac{\Psi^{\mathrm{sc}}_{\ell}(r)}{r}Y_{\ell m}(\theta,\varphi)e^{-i\omega t}.
\end{equation}
The radial function is described by Eq.~\eqref{eq:master_common}, with $\Psi_a=\Psi^{\mathrm{sc}}_{\ell}$ and effective potential
\begin{equation}
\label{eq:scalar_potential}
V^{\mathrm{sc}}_{\ell}(r)=F(r)\frac{\ell(\ell+1)}{r^2}+\frac{1}{2r}\frac{d}{dr}\left[F(r)G(r)\right].
\end{equation}
Using Eq.~\eqref{eq:metric}, this may also be written explicitly as
\begin{equation}
\begin{split}
V^{\mathrm{sc}}_{\ell}(r)&=\left(1-\frac{r_h}{r}\right)\frac{\ell(\ell+1)}{r^2}
+\frac{1}{2r}\frac{d}{dr}\Bigg[
\left(1-\frac{r_h}{r}\right)^2\\
&\hspace{3.5cm}\times
\left(1-\frac{\alpha r_h}{r}\right)\Bigg].
\end{split}
\end{equation}
The holonomy parameter therefore changes both the tortoise stretching and the scalar barrier itself.

\textbf{Electromagnetic field.}
The Maxwell equations,
$
\nabla_{\mu}F^{\mu\nu}=0,
$
can be decomposed into axial and polar sectors. For static spherically symmetric backgrounds these sectors are isospectral, and both reduce to the common master equation, Eq.~\eqref{eq:master_common}, for $\ell\ge1$, with $\Psi_a=\Psi^{\mathrm{em}}_{\ell}$ and potential
\begin{equation}
\label{eq:em_potential}
V^{\mathrm{em}}_{\ell}(r)=F(r)\frac{\ell(\ell+1)}{r^2}.
\end{equation}
In contrast with the scalar case, the electromagnetic potential depends on the holonomy parameter only indirectly through the tortoise coordinate. This will be the source of the different scalar and electromagnetic trends seen below.

\textbf{Dirac field.}
For a massless Dirac field, separation in spinor spherical harmonics leads to a pair of supersymmetric partner potentials. Since the two sectors are isospectral, it is sufficient to work with one representative branch. Writing the reduced radial function as $\Psi^{\mathrm{D}}_{k}(r)$, with spinor-harmonic index $k=1,2,\dots$, we use the common master equation, Eq.~\eqref{eq:master_common}, with $\Psi_a=\Psi^{\mathrm{D}}_{k}$ and potential (see, for instance, \cite{Brill:1957fx, Cho:2003qe, Kanti:2006ua, Jing:2003wq})
\begin{equation}
\label{eq:dirac_potential}
V^{\mathrm{D}}_{k}(r)=W_k(r)^2+\frac{dW_k}{dr_*},
\quad
W_k(r)=\frac{k\sqrt{F(r)}}{r}.
\end{equation}
Equivalently,
\begin{equation}
V^{\mathrm{D}}_{k}(r)=\frac{k^2F(r)}{r^2}+\sqrt{F(r)G(r)}\,\frac{d}{dr}\left(\frac{k\sqrt{F(r)}}{r}\right).
\end{equation}
In this sector the holonomy deformation enters through the tortoise factor and through the $dr_*$-derivative acting on the superpotential. The resulting deformation is milder than in the scalar case, which is why the dominant Dirac transmission threshold changes very little across the range of $\alpha$ considered below.

\textbf{Potential barriers and greybody intuition.}
The effective potentials above provide the simplest way to anticipate the greybody factors. The common master equation, Eq.~\eqref{eq:master_common}, has the form of a one-dimensional scattering problem with $V(r)\to0$ at the horizon and at large radius. A wave with $\omega^2$ well above the barrier is transmitted almost freely, whereas a wave with $\omega^2$ below the barrier top must tunnel. Thus, at fixed frequency, increasing the height or the tortoise-coordinate width of the barrier increases reflection and decreases the greybody factor. This is the qualitative content of the WKB estimate used later in Eq.~\eqref{eq:wkb}.

\begin{figure*}[t]
\centering
\includegraphics[width=0.92\textwidth]{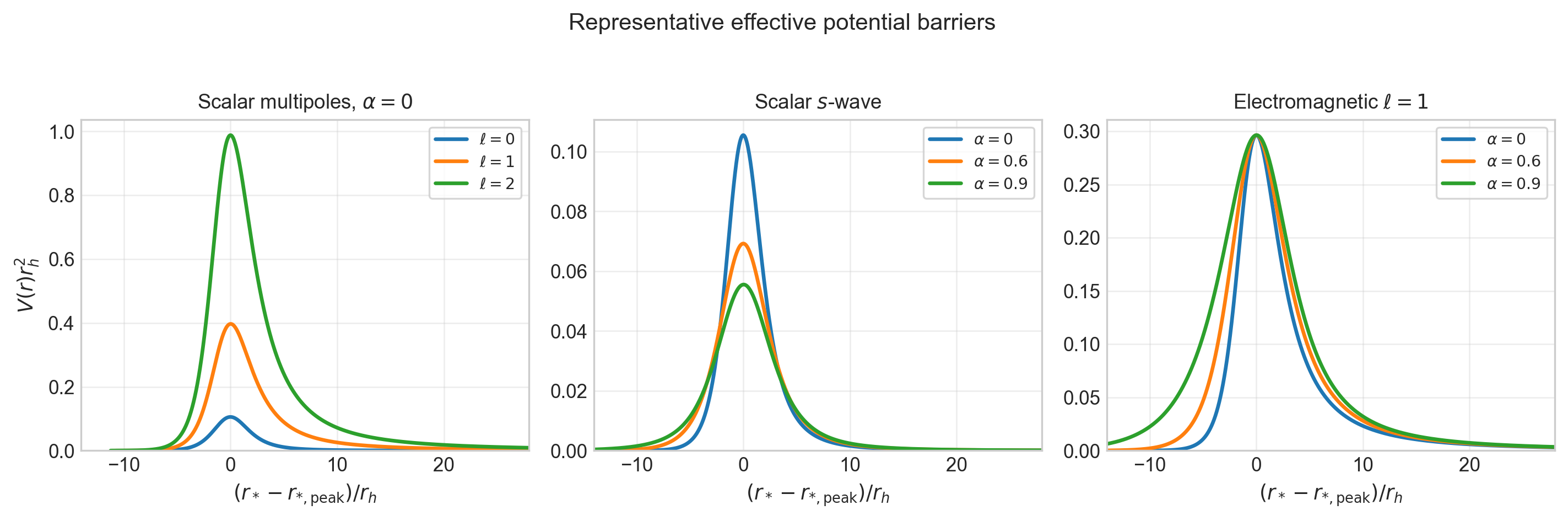}
\caption{Representative effective potential barriers at fixed $r_h=1$, plotted against the tortoise coordinate shifted to the barrier peak. Left: increasing the scalar multipole raises the barrier and therefore suppresses the low-frequency greybody factor. Middle: increasing $\alpha$ lowers the scalar $s$-wave barrier, anticipating enhanced scalar transmission. Right: the electromagnetic $\ell=1$ barrier has the same areal-radius height for every $\alpha$, but the tortoise map broadens it as the holonomy parameter grows, producing a mild suppression of the electromagnetic greybody factor.}
\label{fig:potential_examples}
\end{figure*}

Some explicit examples make this link useful. 
Figure~\ref{fig:potential_examples} displays representative barriers in the shifted tortoise coordinate, so both height and width can be read directly. These examples show two separate effects. First, increasing the angular label raises the centrifugal part of the barrier: for instance, at $\alpha=0$ the scalar peak grows from $V^{\mathrm{sc}}_{0,\max}r_h^2\simeq0.105$ to $V^{\mathrm{sc}}_{1,\max}r_h^2\simeq0.397$. This is why higher multipoles turn on later and have smaller greybody factors in the low-frequency range. Second, the holonomy parameter changes different fields in different ways. The scalar $s$-wave peak decreases from $V^{\mathrm{sc}}_{0,\max}r_h^2\simeq0.105$ at $\alpha=0$ to $0.0555$ at $\alpha=0.9$, which anticipates the enhanced scalar transmission found below. By contrast, $V^{\mathrm{em}}_{1}(r)$ is independent of $\alpha$ as a function of the areal radius, but the tortoise map stretches the barrier as $\alpha$ grows; this broader barrier produces a slightly smaller electromagnetic greybody factor even without a higher areal-radius peak. The Dirac case is intermediate: its barrier is deformed only mildly, so its dominant transmission threshold remains nearly fixed.

\section{Methods}
\label{sec:methods}

\textbf{Direct numerical extraction of greybody factors.}
For each field and multipole we solve the common master equation, Eq.~\eqref{eq:master_common}, with the corresponding scalar, electromagnetic or Dirac potential, by direct numerical integration in the tortoise coordinate. The physical boundary conditions are purely ingoing at the future horizon and a superposition of incoming and outgoing plane waves at infinity:
\begin{equation}
\label{eq:bcs}
\Psi_{\ell}\sim e^{-i\omega r_*}, \quad r_*\to-\infty,
\end{equation}
\begin{equation}
\label{eq:asymptotic}
\Psi_{\ell}\sim A^{\mathrm{in}}_{\ell}(\omega)e^{-i\omega r_*}+A^{\mathrm{out}}_{\ell}(\omega)e^{i\omega r_*},
\quad r_*\to+\infty.
\end{equation}
Near the horizon we start the integration at a small offset from $r_h$ and use the exact ingoing form to set
\begin{equation}
\label{eq:initial_data}
\Psi_{\ell}=1,
\quad
\frac{d\Psi_{\ell}}{dr_*}=-i\omega\Psi_{\ell}.
\end{equation}
At large positive $r_*$ we recover the asymptotic amplitudes from the numerical solution through \cite{Tan:2026vif}
\begin{equation}
\label{eq:amplitudes}
\begin{split}
A^{\mathrm{in}}_{\ell}&=\frac{1}{2}e^{i\omega r_*}
\left(\Psi_{\ell}+\frac{i}{\omega}\frac{d\Psi_{\ell}}{dr_*}\right),\\
A^{\mathrm{out}}_{\ell}&=\frac{1}{2}e^{-i\omega r_*}
\left(\Psi_{\ell}-\frac{i}{\omega}\frac{d\Psi_{\ell}}{dr_*}\right),
\end{split}
\end{equation}
and average the result over the far-zone tail of the solution to reduce residual finite-radius oscillations.

The greybody factor is then
\begin{equation}
\label{eq:gamma_def}
\gamma_{\ell}(\omega)=\left|\mathcal{T}_{\ell}(\omega)\right|^2
=\frac{1}{\left|A^{\mathrm{in}}_{\ell}(\omega)\right|^2}.
\end{equation}
All figures and all quantitative conclusions in this paper are based on these directly integrated transmission probabilities. In the numerical runs reported here the maximum flux imbalance $|\gamma_{\ell}+|\mathcal{R}_{\ell}|^2-1|$ remained below $7.3\times10^{-5}$, which is sufficient for the present comparison.

\begin{figure*}[t]
\centering
\includegraphics[width=\textwidth]{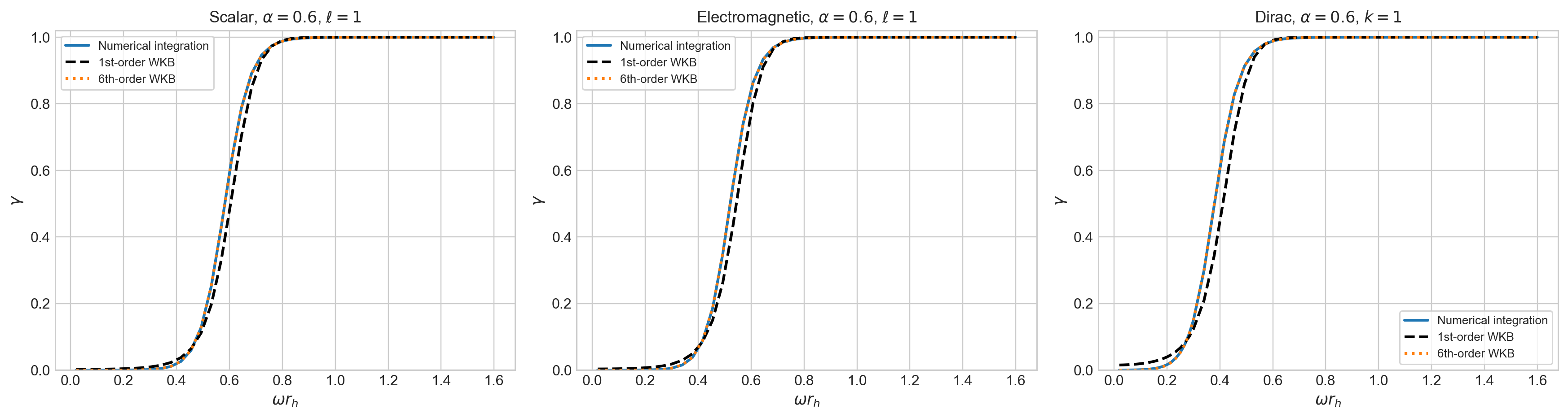}
\caption{Complementary WKB check for the representative modes used in Table~\ref{tab:wkb}: scalar $\ell=1$, electromagnetic $\ell=1$ and Dirac $k=1$, all at $\alpha=0.6$. The dashed curve is the first-order WKB transmission probability of Eq.~\eqref{eq:wkb}, the dotted curve is the sixth-order continued-WKB estimate obtained from the tortoise-coordinate derivatives through $V_0^{(12)}$, and the solid curve is the numerically integrated greybody factor. The sixth-order curve is nearly indistinguishable from the numerical one across the transition region, while the first-order treatment remains only qualitative there.}
\label{fig:wkb}
\end{figure*}

\textbf{Complementary WKB estimate.}
To keep contact with the familiar barrier-top intuition, we compare the numerical transmission curves with the first-order WKB formula of Schutz, Will, Iyer and Will \cite{SchutzWill1985,IyerWill1987}. For a single-peaked barrier with maximum $V_0$ and second tortoise derivative $V_0''<0$, the first-order transmission probability is approximated by
\begin{equation}
\label{eq:wkb}
\gamma^{\mathrm{WKB}}_{\ell}(\omega)\approx
\left[1+\exp\left(2\pi\,\frac{V_0-\omega^2}{\sqrt{-2V_0''}}\right)\right]^{-1}.
\end{equation}
We also evaluate a sixth-order WKB estimate for the same representative modes. This comparison uses the higher-order WKB formula introduced as a practical recipe for quasinormal modes and grey-body factors \cite{Konoplya:2019hlu}. In practice this requires locating the barrier peak, computing tortoise-coordinate derivatives through $V_0^{(12)}$ there, solving the higher-order WKB equation for the barrier parameter $\nu(\omega)$, and reconstructing the transmission probability as $\gamma=[1+\exp(2\pi\nu)]^{-1}$. The same barrier-top logic underlies scalar, Dirac and gravitational perturbation calculations in dilaton, higher curvature, G\"odel, massive-gravity, Einstein--Aether and other backgrounds \cite{Lutfuoglu:2025blw,Bolokhov:2023dxq,Bolokhov:2026dfg,Konoplya:2001ji,Lutfuoglu:2025ldc,Lutfuoglu:2025ljm,Bolokhov:2024ixe,Bolokhov:2024bke,Konoplya:2005sy,Arbelaez:2026eaz,Abdalla:2005hu,Konoplya:2006ar,Fernando:2016ftj,Wongjun:2019ydo,Malik:2024sxv,Malik:2024iky,Konoplya:2024hfg,Lutfuoglu:2026rqe,Lutfuoglu:2025mqa,Lutfuoglu:2025kqp,Lutfuoglu:2025eik,Bolokhov:2026uol,Konoplya:2023ahd,Skvortsova:2026idf,Skvortsova:2026jtx,Skvortsova:2023zmj}. Semi-analytic perturbative treatments in charged Reissner--Nordstr\"om or Reissner--Nordstr\"om--de Sitter backgrounds, magnetized Kerr--Newman spacetimes, squashed Kaluza--Klein black holes, five-dimensional magnetic compact objects and holographic conductivity problems provide additional settings where local approximations must be checked against the relevant scattering or stability problem \cite{Eniceicu:2019npi,Stuchlik:2025mjj,Kokkotas:2010zd,Ishihara:2008re,Bolokhov:2025aqy,Konoplya:2009hv}. Both WKB approximations are used only as consistency checks near the top of the barrier; they are not used in the construction of the greybody, absorption or Hawking-emission plots. Throughout Fig.~\ref{fig:wkb} and Table~\ref{tab:wkb}, the representative channels are the scalar $\ell=1$ mode, the electromagnetic $\ell=1$ mode and the Dirac $k=1$ mode, all at $\alpha=0.6$, so the scalar entry there is not the dominant $\ell=0$ $s$-wave. For each mode we list the three frequencies nearest the numerical half-transmission point. At those benchmark points the first-order WKB error stays between $14.6\%$ and $30.5\%$, whereas the sixth-order error drops to the $0.08\%$--$0.58\%$ range. The comparison confirms the same overall lesson: first-order WKB remains useful for barrier-top intuition, while the sixth-order treatment provides a much sharper semi-analytic cross-check, but the quantitative observables reported in this paper still come from direct numerical transmission probabilities. Higher-order and analytic WKB developments provide useful checks of this kind near the barrier top \cite{Matyjasek:2017psv,Matyjasek:2019eeu,Matyjasek:2026yiu,Konoplya:2023moy}. For greybody factors, however, the usual Pad\'e resummation cannot be applied \cite{Konoplya:2019hlu}; therefore, simply increasing the WKB order does not guarantee a monotonic improvement in accuracy. After all, the WKB method is known to become highly inaccurate, or even inapplicable, in certain theories that produce non-standard centrifugal barriers or effective potentials with multiple peaks and wells, a situation commonly encountered in theories incorporating quantum corrections through higher-curvature terms \cite{Konoplya:2017ymp,Cuyubamba:2016cug,Takahashi:2010gz,Dotti:2004sh}.

\textbf{Absorption cross sections and Hawking emission.}
For the scalar field, the partial and total absorption cross sections are
\begin{equation}
\label{eq:sigma_scalar}
\sigma^{\mathrm{sc}}_{\mathrm{abs}}(\omega)=\sum_{\ell=0}^{\infty}\sigma^{\mathrm{sc}}_{\ell}(\omega),
\quad
\sigma^{\mathrm{sc}}_{\ell}(\omega)=\frac{\pi(2\ell+1)}{\omega^2}\,\gamma^{\mathrm{sc}}_{\ell}(\omega).
\end{equation}
For the electromagnetic field there are two physical parity sectors with the same transmission probability in the test-field Maxwell problem. Averaging over the two incident photon polarizations gives the same equations as for the scalar field.
This normalization is the same as in standard Schwarzschild electromagnetic absorption calculations and makes the high-frequency electromagnetic cross section approach the same geometric-optics capture value as the scalar one \cite{CrispinoOliveiraHiguchiMatsas2007}.

For the Dirac field we use
\begin{equation}
\label{eq:sigma_dirac}
\sigma^{\mathrm{D}}_{\mathrm{abs}}(\omega)=\sum_{k=1}^{\infty}\sigma^{\mathrm{D}}_{k}(\omega),
\quad
\sigma^{\mathrm{D}}_{k}(\omega)=\frac{4\pi k}{\omega^2}\,\gamma^{\mathrm{D}}_{k}(\omega).
\end{equation}

The Hawking energy-emission rates are obtained by summing the thermally weighted greybody factors. For the scalar field, we have
\begin{equation}
\label{eq:hawking_scalar}
\frac{d^2E_{\mathrm{sc}}}{dt\,d\omega}=
\frac{1}{2\pi}
\sum_{\ell=0}^{\infty}(2\ell+1)
\frac{\omega\,\gamma^{\mathrm{sc}}_{\ell}(\omega)}{e^{\omega/T_H}-1}.
\end{equation}
The same rates are obtained for the electromagnetic field once we sum the two physical polarizations.
For the Dirac field the Fermi--Dirac factor and spinor multiplicity give
\begin{equation}
\label{eq:hawking_dirac}
\frac{d^2E_{\mathrm{D}}}{dt\,d\omega}=
\frac{1}{2\pi}
\sum_{k=1}^{\infty}4k
\frac{\omega\,\gamma^{\mathrm{D}}_{k}(\omega)}{e^{\omega/T_H}+1}.
\end{equation}
In addition to the single-field spectra, we also display the Page-style aggregate \cite{Page1976}
\begin{equation}
\label{eq:hawking_page}
\frac{d^2E_{\mathrm{Page}}}{dt\,d\omega}=3\,\frac{d^2E_{\mathrm{D}}}{dt\,d\omega}+\frac{d^2E_{\mathrm{em}}}{dt\,d\omega},
\end{equation}
which serves only as a compact radiative diagnostic rather than as a complete Standard Model particle-physics inventory. We truncate the sums at $\ell_{\max}=6$ for the scalar field, $\ell_{\max}=7$ for the electromagnetic field and $k_{\max}=7$ for the Dirac field. In the frequency region that controls the Hawking peaks, the highest included multipole is already negligible on the scale of the plots.

Equation~\eqref{eq:hawking_scalar} is used in the usual fixed-background, semiclassical sense. This is appropriate here because the energy carried by a single emitted quantum is much smaller than the black-hole mass in the regime we analyze, so the geometry changes only adiabatically during a single emission event.

\section{Greybody Factors and Absorption Cross Sections}
\label{sec:gbf}

\begin{figure*}[t]
\centering
\includegraphics[width=0.75\textwidth]{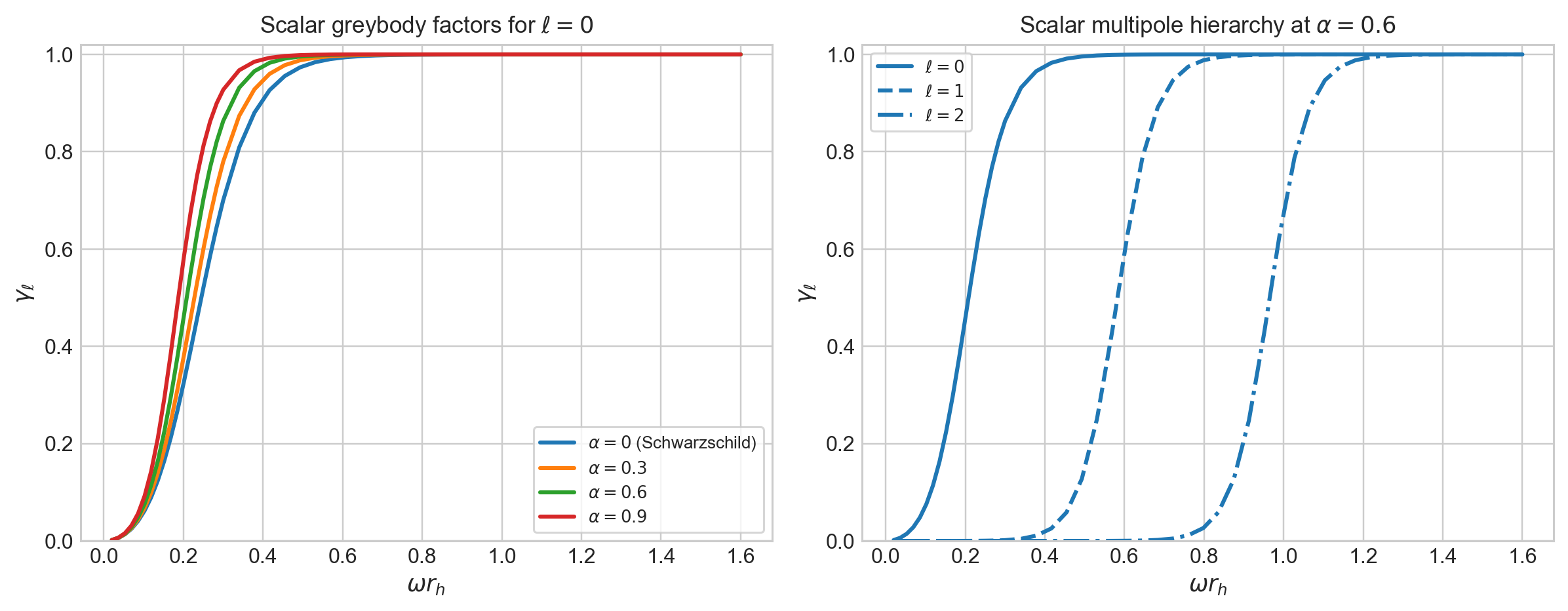}
\caption{Scalar greybody factors. Left: dependence of the dominant $\ell=0$ mode on the holonomy parameter. Right: scalar multipole hierarchy at the representative value $\alpha=0.6$.}
\label{fig:scalar_gbf}
\end{figure*}

\textbf{Greybody Factors.} We first discuss the transmission coefficients themselves. Figure~\ref{fig:scalar_gbf} shows the scalar greybody factors. The left panel isolates the dominant $s$-wave, which is the mode most relevant for low-frequency scattering and Hawking emission. As the holonomy parameter increases from $\alpha=0$ to $0.9$, the curve moves steadily to the left: the half-transmission point decreases from $\omega r_h\simeq0.245$ to $0.187$. In other words, the holonomy correction makes the dominant scalar mode more transparent.

This behavior is easy to understand qualitatively from Eq.~\eqref{eq:scalar_potential}. The correction alters not only the tortoise stretching but also the barrier itself through the derivative term $\frac{1}{2r}d(FG)/dr$. For the lowest scalar modes this lowers the effective threshold sufficiently to compensate the broadening of the tortoise coordinate. The right panel of Fig.~\ref{fig:scalar_gbf} shows that the standard multipole hierarchy is preserved: higher $\ell$ still switch on at larger frequencies, but the scalar spectrum remains smoothly ordered and numerically well behaved.

\begin{figure*}[t]
\centering
\includegraphics[width=0.8\textwidth]{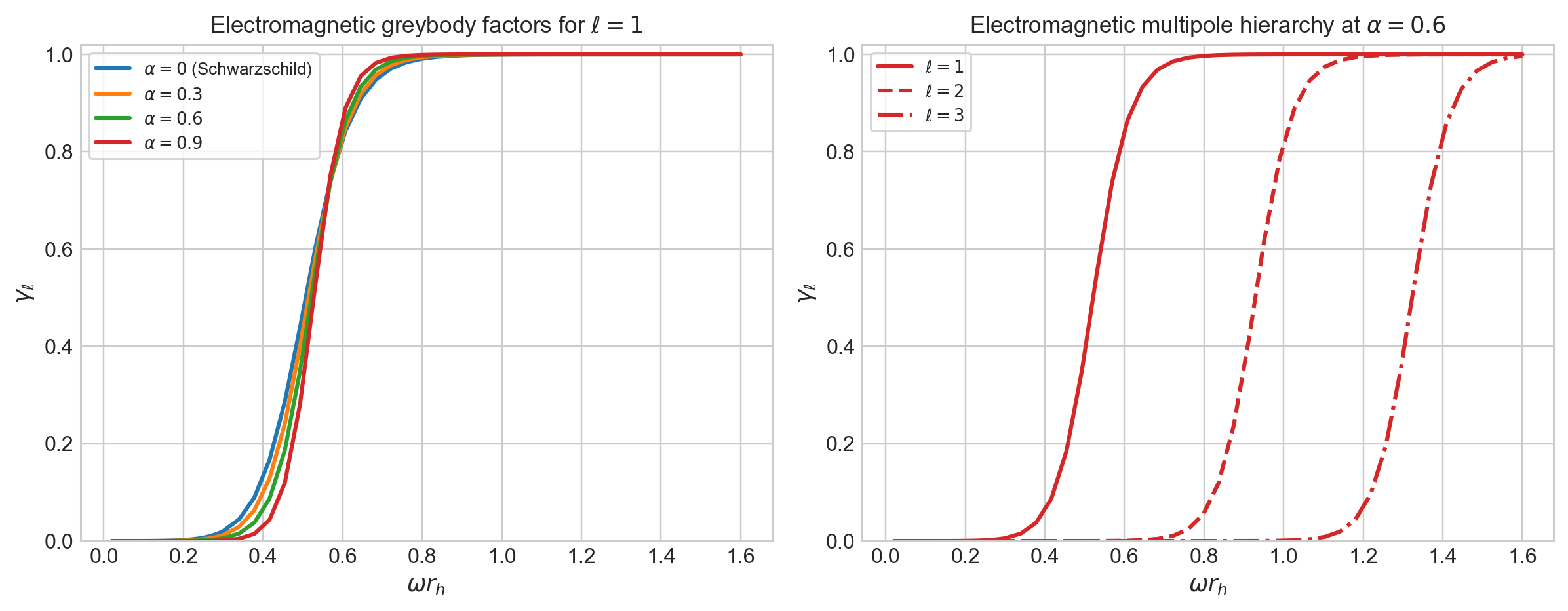}
\caption{Electromagnetic greybody factors. Left: dependence of the dominant electromagnetic mode $\ell=1$ on the holonomy parameter. Right: multipole hierarchy at $\alpha=0.6$.}
\label{fig:em_gbf}
\end{figure*}

The electromagnetic sector, displayed in Fig.~\ref{fig:em_gbf}, reacts differently. The left panel shows that the $\ell=1$ transmission curve shifts slightly to the right as $\alpha$ grows, with the half-transmission point moving from $\omega r_h\simeq0.507$ to $0.527$. This is a small effect, but it is systematic. Since the electromagnetic potential \eqref{eq:em_potential} retains its Schwarzschild form in the areal radius, the holonomy parameter enters only through the tortoise map, which stretches the barrier in $r_*$. The result is a mildly reduced electromagnetic transparency. The right panel of Fig.~\ref{fig:em_gbf} again shows the expected suppression of higher multipoles.

\begin{figure*}[t]
\centering
\includegraphics[width=0.75\textwidth]{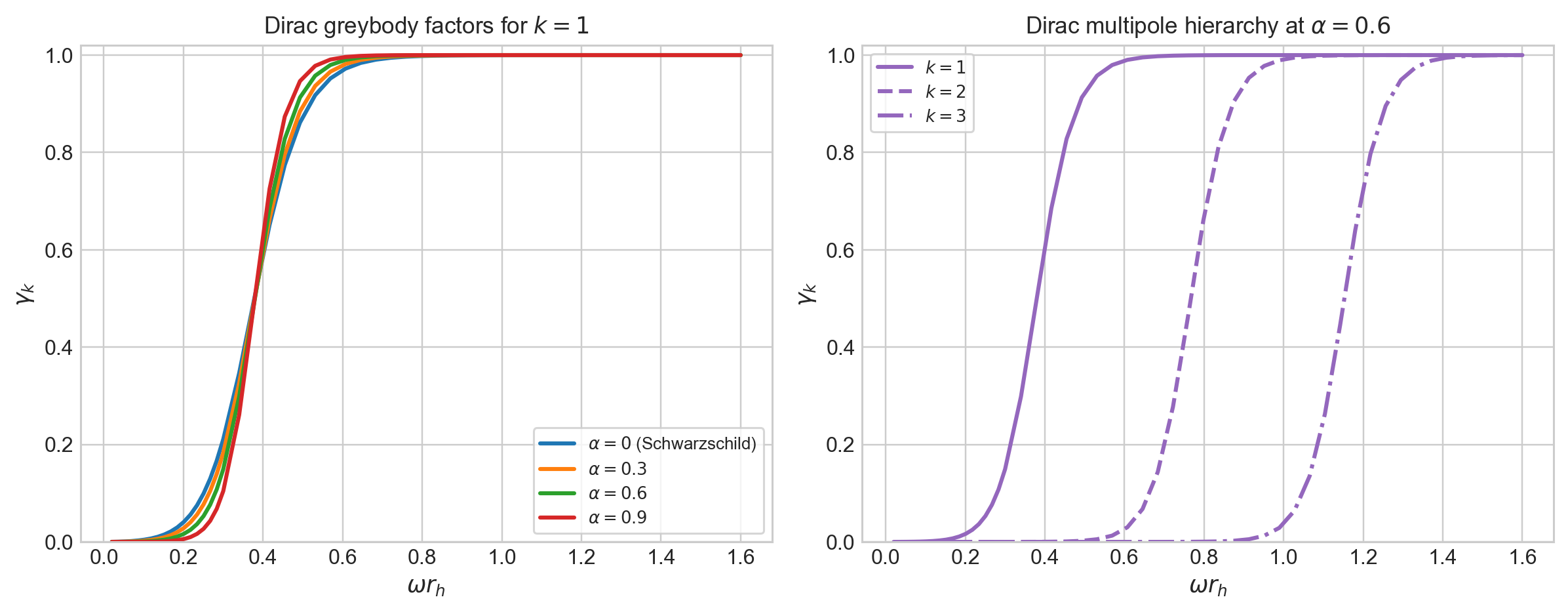}
\caption{Dirac greybody factors. Left: dependence of the dominant Dirac mode $k=1$ on the holonomy parameter. Right: multipole hierarchy at $\alpha=0.6$.}
\label{fig:dirac_gbf}
\end{figure*}

The Dirac sector, shown in Fig.~\ref{fig:dirac_gbf}, displays a third pattern. The dominant $k=1$ mode is remarkably stable under the holonomy deformation: its half-transmission point stays essentially fixed, moving only from $\omega r_h\simeq0.378$ at $\alpha=0$ to $0.379$ at $\alpha=0.9$. Thus the dominant fermionic threshold is neither enhanced as strongly as the scalar one nor pushed upward as in the electromagnetic case. The right panel shows that the usual hierarchy with increasing $k$ is still present, so the holonomy correction preserves the basic partial-wave ordering while only weakly deforming the lowest Dirac channel.

The contrast among the three sectors is one of the central qualitative findings of the paper. The same geometric deformation that makes the scalar $s$-wave easier to transmit makes the electromagnetic barrier slightly harder to cross, while leaving the dominant Dirac threshold almost unchanged. This matters for absorption, but it becomes even more important once thermal emission is included.

\begin{table*}[t]
\centering
\caption{Representative WKB-versus-numerical comparison points from Fig.~\ref{fig:wkb}. The table refers to the scalar $\ell=1$ mode, the electromagnetic $\ell=1$ mode and the Dirac $k=1$ mode at $\alpha=0.6$; in particular, the scalar sample here is not the $\ell=0$ mode. For each channel we list the three frequencies nearest the numerical half-transmission point. The frequency column is the dimensionless product $\omega r_h$, all greybody factors $\gamma$ are dimensionless transmission probabilities, and the error columns are percentages. The two WKB columns are the first-order result $\gamma^{(1)}_{\mathrm{WKB}}$ from Eq.~\eqref{eq:wkb} and the sixth-order continued-WKB value $\gamma^{(6)}_{\mathrm{WKB}}$ computed from the same barrier peak with tortoise-coordinate derivatives through $V_0^{(12)}$. The last two columns are $100\,|\gamma^{(1)}_{\mathrm{WKB}}-\gamma_{\mathrm{num}}|/\gamma_{\mathrm{num}}$ and $100\,|\gamma^{(6)}_{\mathrm{WKB}}-\gamma_{\mathrm{num}}|/\gamma_{\mathrm{num}}$, respectively. The first-order error stays at the $15$--$31\%$ level near the transition region, while the sixth-order result is uniformly sub-percent.}
\label{tab:wkb}
\small
\begin{tabular}{cccccccc}
\toprule
Field & mode & $\omega r_h$ (dimensionless) & $\gamma_{\mathrm{num}}$ (dimensionless) & $\gamma^{(1)}_{\mathrm{WKB}}$ (dimensionless) & $\gamma^{(6)}_{\mathrm{WKB}}$ (dimensionless) & err.$^{(1)}$ (\%) & err.$^{(6)}$ (\%) \\
\midrule
        Scalar & $\ell=1$ & 0.531 & 0.250 & 0.195 & 0.252 & 22.00\% & 0.43\% \\
        Scalar & $\ell=1$ & 0.569 & 0.431 & 0.331 & 0.431 & 23.15\% & 0.11\% \\
        Scalar & $\ell=1$ & 0.607 & 0.628 & 0.514 & 0.628 & 18.19\% & 0.10\% \\
        EM & $\ell=1$ & 0.493 & 0.349 & 0.262 & 0.350 & 24.85\% & 0.46\% \\
        EM & $\ell=1$ & 0.531 & 0.553 & 0.431 & 0.553 & 22.10\% & 0.11\% \\
        EM & $\ell=1$ & 0.569 & 0.738 & 0.630 & 0.737 & 14.59\% & 0.08\% \\
        Dirac & $k=1$ & 0.340 & 0.299 & 0.209 & 0.301 & 30.00\% & 0.58\% \\
        Dirac & $k=1$ & 0.378 & 0.493 & 0.342 & 0.495 & 30.54\% & 0.50\% \\
        Dirac & $k=1$ & 0.416 & 0.686 & 0.524 & 0.687 & 23.54\% & 0.26\% \\
        \bottomrule
\end{tabular}
\end{table*}

\begin{figure*}[t]
\centering
\includegraphics[width=\textwidth]{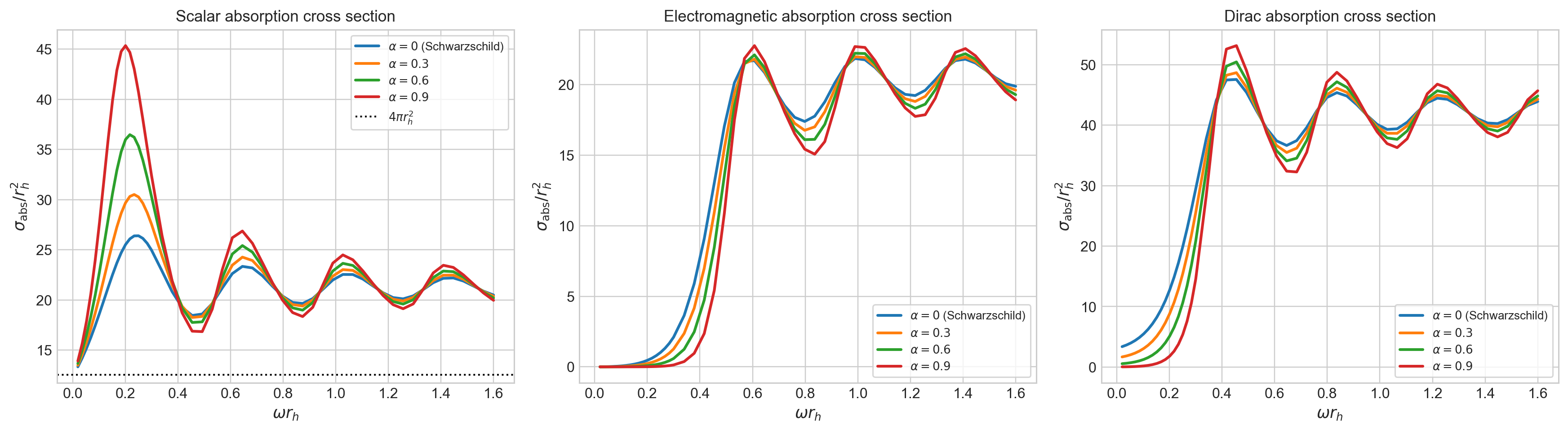}
\caption{Total absorption cross sections for scalar, electromagnetic and Dirac test fields. The dotted horizontal line marks the scalar low-frequency limit $4\pi r_h^2$. The scalar channel is strongly enhanced at the first oscillation peak as $\alpha$ increases, the electromagnetic cross section mainly changes in the infrared before approaching a common high-frequency envelope, and the Dirac cross section develops a strongly suppressed low-frequency tail while keeping a broad peak at nearly fixed frequency.}
\label{fig:absorption}
\end{figure*}

\textbf{Absorption Cross Sections.}  The total absorption cross sections are shown in Fig.~\ref{fig:absorption}. For the scalar field, the low-frequency limit is consistent with the universal result $\sigma_{\mathrm{abs}}\to 4\pi r_h^2$ for minimally coupled massless scalars \cite{DasGibbonsMathur1997}. Numerically, the first plotted point at $\omega r_h=0.02$ already lies close to this limit for every $\alpha$, and the remaining discrepancy is simply the fact that the plot starts at a small but nonzero frequency.

Beyond the threshold region, however, the holonomy correction has a strong effect on the scalar absorption pattern. The first oscillation peak grows from $\sigma_{\mathrm{abs}}^{\mathrm{sc}}/r_h^2\simeq26.4$ at $\omega r_h\simeq0.251$ for Schwarzschild to $45.4$ at $\omega r_h\simeq0.201$ for $\alpha=0.9$. Thus the holonomy correction not only enhances the dominant scalar transmission, but also shifts the main absorption structure to lower frequencies while increasing its amplitude.

The electromagnetic absorption cross section behaves differently. At very low frequencies the absorption is strongly suppressed as $\alpha$ grows, which mirrors the rightward shift of the electromagnetic greybody factors. At intermediate and high frequencies, however, the total cross sections cluster around the same geometric-optics envelope, with peaks near $\sigma_{\mathrm{abs}}^{\mathrm{em}}/r_h^2\approx 22$. This is exactly what one should expect once the transmission coefficients are close to unity, the remaining structure is governed mainly by the usual partial-wave interference pattern, and the holonomy correction survives only as a mild phase-level deformation.

The Dirac absorption cross section defines a third pattern. Its deep-infrared tail is strongly damped by the holonomy correction: the first plotted point decreases from $\sigma_{\mathrm{abs}}^{\mathrm{D}}/r_h^2\simeq3.38$ at $\alpha=0$ to $3.72\times10^{-2}$ at $\alpha=0.9$. At the same time, the broad maximum grows only mildly, from about $47.6$ to $53.2$, and stays near $\omega r_h\simeq0.455$. In other words, the deformation suppresses low-energy fermionic absorption strongly while leaving the main Dirac resonance scale nearly fixed.

In short, the absorption problem separates into three distinct stories. The scalar cross section is significantly amplified by the holonomy correction in the low and intermediate frequency regimes, the electromagnetic cross section mainly feels the correction through a suppression of the deep infrared region while preserving a near-Schwarzschild high-frequency envelope, and the Dirac cross section combines a strong infrared damping with only a mild change in its main peak structure.

\section{Hawking Radiation}
\label{sec:hawking}

\begin{figure*}[t]
\centering
\includegraphics[width=0.75\textwidth]{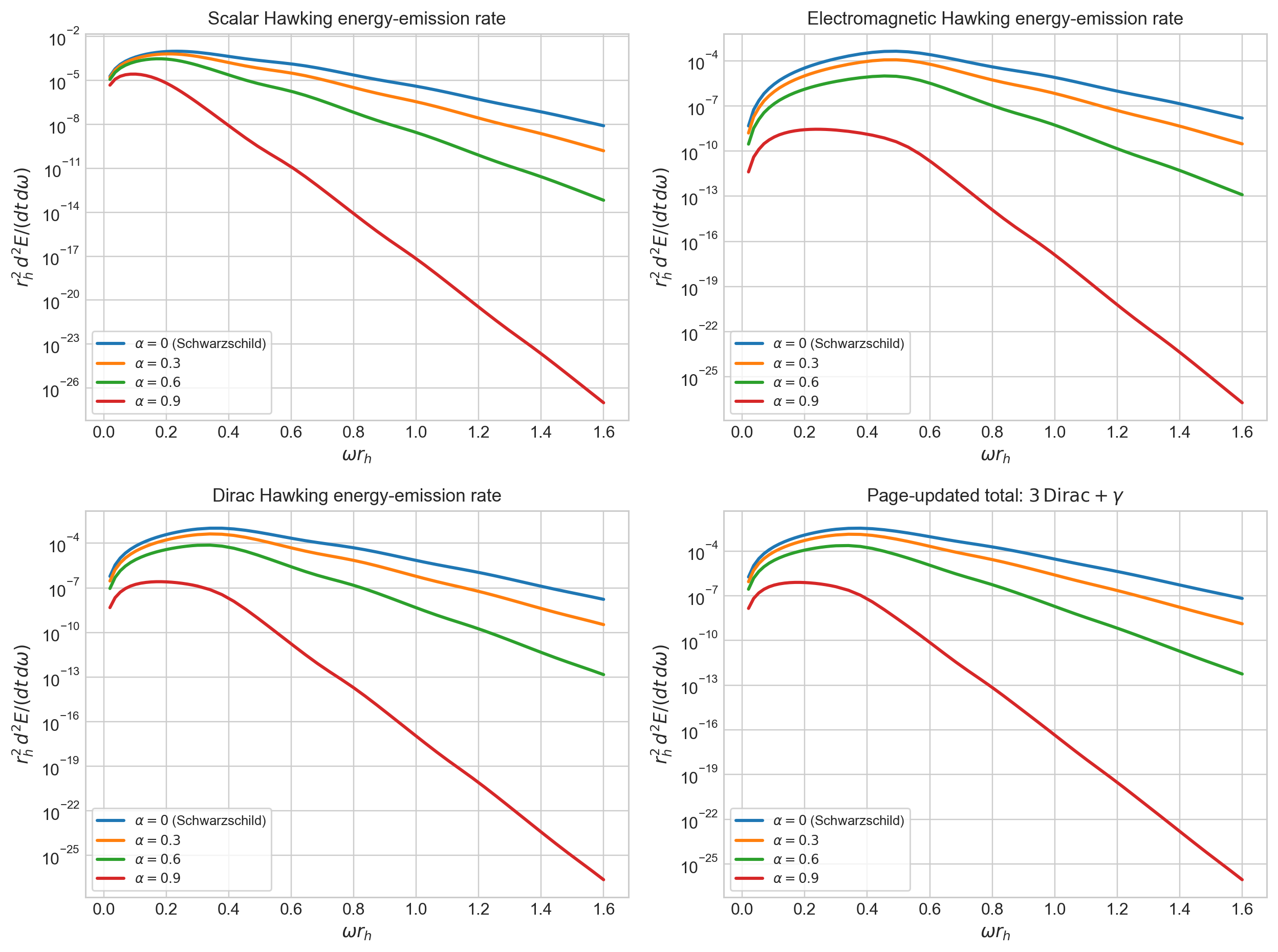}
\caption{Hawking energy-emission rates. The four panels show the scalar, electromagnetic, Dirac and Page-style diagnostic spectra. In every case the spectrum shifts to lower frequencies and is strongly suppressed as $\alpha$ increases. The electromagnetic channel is quenched most efficiently, while the Page-style diagnostic becomes Dirac-dominated once $\alpha$ is appreciable.}
\label{fig:emission}
\end{figure*}

The Hawking energy-emission rates are displayed in Fig.~\ref{fig:emission}, and the corresponding integrated powers are summarized in Table~\ref{tab:power}. This is where the cooling effect of Eq.~\eqref{eq:temperature} becomes decisive.

For the scalar field, the enhanced low-lying transmission does \emph{not} translate into a stronger emitted flux. Instead, the thermal factor overwhelms the gain in transparency. The peak of the scalar emissivity moves from $\omega r_h\simeq0.234$ at $\alpha=0$ to $0.102$ at $\alpha=0.9$, while the integrated scalar power drops from $2.97\times10^{-4}$ to $3.46\times10^{-6}$. That is a suppression by roughly a factor of $8.6\times10^{1}$ despite the fact that the scalar $s$-wave itself is easier to transmit.

The electromagnetic channel is suppressed even more strongly. Here the holonomy correction acts in the same direction on both ingredients entering the Hawking flux: the transmission becomes slightly less favorable and the temperature becomes much smaller. The peak of the electromagnetic emissivity slides from $\omega r_h\simeq0.493$ to $0.234$, but the integrated power collapses from $1.34\times10^{-4}$ to $7.56\times10^{-10}$ as $\alpha$ grows from $0$ to $0.9$. This is a reduction by about $5.6\times10^{-6}$.

The Dirac channel is also strongly quenched, even though its dominant transmission threshold barely changes. The peak of the fermionic emissivity moves from $\omega r_h\simeq0.378$ at $\alpha=0$ to $0.185$ at $\alpha=0.9$, and the integrated power decreases from $3.26\times10^{-4}$ to $5.71\times10^{-8}$. This is a suppression by roughly a factor of $5.7\times10^{3}$. In the Dirac sector the reduction is therefore driven overwhelmingly by cooling and by the Fermi--Dirac occupation factor, not by any major shift in the dominant transmission threshold itself.

The bottom-right panel of Fig.~\ref{fig:emission} packages these results into the Page-aggregate $3\,\mathrm{Dirac}+\gamma$. The corresponding integrated aggregate power drops from $1.11\times10^{-3}$ to $1.72\times10^{-7}$ across the same range of $\alpha$ and is quickly dominated by the Dirac contribution because the electromagnetic channel is quenched even more aggressively.

Table~\ref{tab:power} organizes the integrated radiative output in a compact way. The scalar half-transmission point decreases with $\alpha$, confirming the increased scalar transparency. The electromagnetic half-transmission point increases slightly, confirming the opposite trend in that sector. The Dirac half-transmission point, by contrast, stays essentially pinned at $\omega r_h\simeq0.379$. Yet all integrated powers fall rapidly because $T_H\propto\sqrt{1-\alpha}$. The scalar field therefore provides the cleanest example of the central message, while the Dirac sector shows the complementary lesson: even an almost unchanged transmission threshold does not prevent Hawking radiation from being strongly suppressed when the black hole becomes colder.

The Schwarzschild limit also provides a normalization check against Page's calculation for an uncharged, nonrotating black hole \cite{Page1976}. Page quotes rates in mass units, whereas Table~\ref{tab:power} uses fixed-$r_h$ units. Since $r_h=2M$, the conversion is $M^2\mathcal{P}=(r_h^2\mathcal{P})/4$. Applying this to the $\alpha=0$ row gives $M^2\mathcal{P}_{\mathrm{sc}}=7.42\times10^{-5}$, $M^2\mathcal{P}_{\mathrm{em}}=3.35\times10^{-5}$ and $M^2\mathcal{P}_{\mathrm{D}}=8.16\times10^{-5}$. These values match the Schwarzschild scalar and photon benchmarks and are consistent with Page's spin-$1/2$ result once the Dirac-field multiplicity used in Eq.~\eqref{eq:hawking_dirac} is accounted for. The table's $3\,\mathrm{Dirac}+\gamma$ column should therefore be read only as a Page-style diagnostic with three Dirac fields and photons, not as Page's complete massless-particle inventory, which also included a graviton contribution. Relative to the Schwarzschild values in the table, the $\alpha=0.9$ scalar, electromagnetic, Dirac and $3\,\mathrm{Dirac}+\gamma$ powers are suppressed to $1.17\times10^{-2}$, $5.64\times10^{-6}$, $1.75\times10^{-4}$ and $1.55\times10^{-4}$, respectively.

\begin{table*}[t]
\centering
\caption{Representative thermal and integrated-power diagnostics for the holonomy-corrected black hole at fixed $r_h=1$. The holonomy parameter $\alpha$ is dimensionless; $T_H$ is quoted in units of $r_h^{-1}$; and the integrated powers are quoted in units of $r_h^{-2}$, so the displayed numbers are the dimensionless coefficients obtained after setting $r_h=1$. The integrated powers are computed from the numerically determined spectra with truncated sums $\ell\le6$ (scalar), $\ell\le7$ (electromagnetic) and $k\le7$ (Dirac). The last column is the Page-style diagnostic aggregate $3\,\mathrm{Dirac}+\gamma$.}
\label{tab:power}
\begin{tabular}{cccccc}
\toprule
$\alpha$ (dimensionless) & $T_H$ [$r_h^{-1}$] & $\mathcal{P}_{\mathrm{sc}}$ [$r_h^{-2}$] & $\mathcal{P}_{\mathrm{em}}$ [$r_h^{-2}$] & $\mathcal{P}_{\mathrm{D}}$ [$r_h^{-2}$] & $\mathcal{P}_{\mathrm{Page}}$ [$r_h^{-2}$] \\
\midrule
        0.0 & 0.0796 & $2.968\times10^{-4}$ & $1.339\times10^{-4}$ & $3.262\times10^{-4}$ & $1.113\times10^{-3}$ \\
        0.3 & 0.0666 & $1.615\times10^{-4}$ & $3.368\times10^{-5}$ & $1.183\times10^{-4}$ & $3.886\times10^{-4}$ \\
        0.6 & 0.0503 & $5.999\times10^{-5}$ & $2.671\times10^{-6}$ & $1.908\times10^{-5}$ & $5.990\times10^{-5}$ \\
        0.9 & 0.0252 & $3.461\times10^{-6}$ & $7.558\times10^{-10}$ & $5.713\times10^{-8}$ & $1.721\times10^{-7}$ \\
\bottomrule
\end{tabular}
\end{table*}

\section{Conclusion}
\label{sec:conclusion}

We have constructed a focused greybody-factor analysis for the holonomy-corrected Schwarzschild black hole, covering the full chain from the effective metric to scattering, absorption and Hawking radiation for scalar, electromagnetic and Dirac test fields. The results are simple to state but physically revealing.

First, the holonomy correction does not affect all test fields in the same way. The scalar field, especially its dominant $\ell=0$ mode, becomes more transparent as $\alpha$ increases. By contrast, the electromagnetic field becomes slightly less transparent because its potential keeps the Schwarzschild form in the areal radius while the tortoise coordinate stretches the barrier. The dominant Dirac mode sits between these behaviors: its transmission threshold is almost unchanged, although the full fermionic spectrum is still deformed. The geometry therefore splits scalar, electromagnetic and fermionic transmission already at the level of the effective one-dimensional problem.

Second, the absorption cross sections make this split visible in a directly observable quantity. The scalar cross section retains the expected low-frequency limit $4\pi r_h^2$ and develops increasingly pronounced low-frequency peaks as the holonomy parameter grows. The electromagnetic cross section mainly changes in the infrared and then returns to a nearly common high-frequency envelope. The Dirac cross section is strongly damped in the deep infrared, yet its main broad maximum stays close to the same frequency and increases only mildly.

Third, and most importantly, Hawking radiation is controlled by the competition between transmission and temperature. Because the holonomy correction lowers the temperature as $\sqrt{1-\alpha}$, scalar, electromagnetic and Dirac emission are all suppressed overall. The suppression is moderate but unmistakable for the scalar field, dramatic for the electromagnetic field, and strong even for the Dirac field despite its nearly unchanged dominant threshold. In that sense the holonomy-corrected black hole is best described as a colder emitter whose scalar $s$-wave is nevertheless more transparent than in Schwarzschild and whose Page-style total rapidly becomes fermion dominated.

This paper stays deliberately within the clean test-field regime, so there are several natural extensions. One may include massive fields, nonminimal couplings, or the sparsity diagnostics of the Hawking cascade. It would also be interesting to combine the present exact transmission analysis with the quasinormal-mode program already developed for this geometry, and to push the multipole sums further into the high-frequency regime where the geometric-optics limit can be studied more precisely. Long-lived modes and overtone behavior in Bardeen, regular or extreme, effective-quantum, quantum Oppenheimer--Snyder and holonomy-corrected black holes form a natural comparison set for such a combined transmission--ringing study \cite{Bolokhov:2023ruj,Skvortsova:2024eqi,Bolokhov:2024bke,Skvortsova:2024atk,Skvortsova:2024wly,Bolokhov:2023bwm}. Another natural extension would be to test the robustness of the present greybody factors and Hawking spectra under small localized static deformations of the holonomy-corrected exterior, in the same spirit as the near-horizon and environmental deformations considered in Ref.~\cite{Konoplya:2025ixm}. But even at the present level, the overall message is already clear: in the holonomy-corrected black hole, transmission and thermodynamics are deformed in different ways for scalar, electromagnetic and fermionic probes, and their competition leaves a sharp imprint on absorption and Hawking radiation.

\section*{Declaration of Competing Interest}
The authors declare that they have no known competing financial interests or personal relationships that could have appeared to influence the work reported in this paper.

\section*{Data Availability}
No data was used for the research described in the article.

\begin{acknowledgments}
B. C. L. is grateful to the Excellence project FoS UHK 2205/2025-2026 for the financial support.
\end{acknowledgments}

\bibliographystyle{apsrev4-1}
\bibliography{holonomy_corrected_gbf}

\end{document}